\documentclass[preprint,12pt]{elsarticle}
\usepackage{lineno,hyperref}
\modulolinenumbers[5]

\usepackage{amssymb}

\usepackage{graphicx}
\usepackage{amsmath}
\usepackage{slashed}

\usepackage{xcolor}

\journal{ Physics Letters B}

\begin{document}

\begin{frontmatter}



\title{Searching for Heavy Neutrinos with the MoEDAL-MAPP Detector at the LHC}


\author{Mariana Frank$^a$}
\address[1]{Department of Physics, Concordia University, 
  Montreal, Quebec, H4B 1R6, Canada}
\author{Marc de Montigny$^b$\tnoteref{correspondingauthor}}
\tnotetext[correspondingauthor]{Corresponding Author }
\address[b]{Facult\'e Saint-Jean, University of Alberta, 
  Edmonton, Alberta, T6C 4G9, Canada}
\author{Pierre-Philippe A. Ouimet$^c$}
\address[c]{Department of Physics, University of Regina, 
 Regina, Saskatchewan, S4S 0A2, Canada}
  \author{James Pinfold$^d$}
\address[d]{Department of Physics, University of Alberta, 
Edmonton, Alberta, T6G 2E1, Canada}
  \author{Ameir Shaa$^d$}
  \author{Michael Staelens$^d$}


\begin{abstract}
We present a  strategy for searching for heavy neutrinos at the Large Hadron Collider using the MoEDAL Experiment's MAPP detector. We hypothesize  the heavy neutrino to be  a member of a fourth generation lepton doublet, with the electric dipole moment (EDM) introduced within a dimension-five operator. In this model the heavy neutrino is  produced in association with a heavy lepton. According to our current experimental and theoretical understanding,  the electric dipole moment of this heavy neutrino may be as high as $10^{-15}$ $e$ cm.  Taking advantage of the sensitivity of MoEDAL detector, we  
examine the possibility of detecting such a heavy neutrino in the MAPP as an apparently  fractionally charged particle, via ionization due to the neutrino's EDM. 
\end{abstract}



\begin{keyword}
Heavy neutrino, electric dipole moment, heavy lepton, LHC, MilliQan,  MoEDAL



\end{keyword}

\end{frontmatter}


\section{Introduction}
\label{introduction}

In this work, we wish to explore the possibility that a heavy neutrino  with a large electric dipole moment would be detectable by MoEDAL's (Monopole and Exotics Detector at the LHC) MAPP (MoEDAL Apparatus for Penetrating Particles) subdetector. MoEDAL is the seventh and newest experiment at the Large Hadron Collider (LHC)  \cite{MoEDAL, Design,Fairbairn:2016usa}. MAPP will be installed adjacent to the MoEDAL detectors in order to take data during Run-3 of the LHC.  

We consider the possibility here that the heavy neutrino could be revealed at the LHC via the ionization caused by an anomalously large electric dipole moment of the neutrino.  There are several possible models that could possibly give rise to such large electric dipole moments (EDMs). In order to cast as  wide a net as possible we utilized an  effective Lagrangian approach to modelling heavy neutrinos with sizeable EDM.  The possibility of electric dipole moment (EDM) of a heavy neutrino was discussed previously in Refs. \cite{Sher:2001rk,Sher:2002ij,Sher:2017wya}; earlier calculations of electromagnetic properties of neutrinos are in Ref. \cite{Shrock:1982sc}. 

The detection of permanent EDMs of particle would provide incontrovertible evidence of physics beyond the SM. As EDMs violate both parity and time-reversal symmetries, their measurement would allow the further elucidation  of CP-violation at the TeV scale. Current complementary experiments to those done at the LHC  are designed to be sensitive to the supersymmetry range of EDMs \cite{Ibrahim:2014oia}.

In the Standard Model (SM), the EDMs are exceedingly small \cite{Hoogeveen:1990cb,Pospelov:1991zt}. The ACME (Advanced Cold Molecule Electron EDM) experiment \cite{Baron:2013eja} improved the previous bound \cite{DeMille:1994qr,Abdullah:1990nh} on the electron EDM of $4.3\times 10^{-27}\ e$ cm to $0.87\times 10^{-28}\ e$ cm. More recent results provide slightly tighter upper bounds of $\left|d_E\right|<1.3\times 10^{-28}\ e$ cm \cite{Cairncross:2017fip} and $\left|d_E\right|<9.4\times 10^{-29}\ e$ cm \cite{Baron:2016obh}. The Particle Data Group now gives an electron EDM limit of $\left|d_E\right|< 0.11 \times 10^{-28} \ e$ cm \cite{Tanabashi:2018oca}. The muon EDM limit was $1.1\times10^{-18}\ e$ cm in 1978 \cite{Bailey:1977sw}, and was lowered thirty years later to $\left|d_\mu\right|<1.8\times 10^{-19}\ e$ cm in Ref. \cite{Bennett:2008dy}. For the muon, the Particle Data Group gives a limit of $\left|d_\mu\right|<-0.1 \pm 0.9 \times 10^{-19}\ e$ cm \cite{Tanabashi:2018oca}. The limit on the tau's EDM was listed at $3\times 10^{-16}\ e$ cm in 2001 \cite{Groom:2000in} and corrected to $-2.2\times10^{-17}<Re(d_\tau)<4.5\times 10 ^{-17}\ e$ cm in Ref. \cite{Inami:2002ah}. 

From a theoretical perspective, values of the EDM are model dependent. For instance, in multiple-Higgs models, the EDM of the muon are at most as $10^{-24}\ e$ cm. (For a detailed discussion of EDMs in multiple Higgs models, see Ref. \cite{Barger:1996jc}.) In leptoquark models, the muon and tau EDMs are again typically $10^{-24}\ e$ cm and $10^{-19}\ e$ cm, respectively \cite{Bernreuther:1996dr}. Likewise, in left-right models \cite{Nieves:1986uk,Valle:1983nx,Cheng:1986in,Babu:2000dq}, the muon EDM is typically ($10^{-24}\ e$ cm) $\sin\alpha$, where $\alpha$ is a phase angle. Additionally,  in the minimal supersymmetric standard model (MSSM) \cite{delAguila:1983dfr}, the electron EDM is somewhat above the experimental bounds if the phases are all of order unity. 

Babu, Barr, and Dorsner \cite{Babu:2000cz} discussed how the EDMs of leptons scale with the lepton masses. In many models, such as the MSSM, they scale linearly with the mass. However, in a number of models, such as some multiple-Higgs, leptoquark, and flavor symmetry models, the EDM scale as the cube of the lepton mass. In these models the tau EDM will be $5000$ times larger than the muon EDM. More details can be found in a  theoretical review of EDM beyond the  SM in Ref. \cite{Fukuyama:2012np}, with further clarifications in Ref. \cite{Fukuyama:2015yya}. From this we see that a wide variety of models with new  heavy leptons give rise to  EDMs that may  be observable in the next round of experiments. As was done in Refs. \cite{Sher:2001rk,Sher:2002ij,Sher:2017wya} we will adopt an upper bound of $10^{-15}\ e $ cm for our EDM in this work.

Current experimental bounds on heavy neutral leptons require that the mass of the heavy neutrino be larger than 45 GeV \cite{Tanabashi:2018oca}. Upper bounds on possible neutrino masses are model dependent. If the heavy neutrino is part of a fourth generation of fermions then it cannot be accommodated by a minimal extension of the SM as this is ruled out by Higgs data, in particular the $H \rightarrow \gamma \gamma$ decay \cite{Giardino:2013bma}. However, other models do allow for a fourth generation of fermions (vector-like) \cite{Ishiwata:2011hr} and therefore for a heavy neutrino that is a member of a fourth leptonic $SU(2)$ isodoublet. Its heavy charged partner would then need to have a mass greater than 100.8 GeV \cite{Tanabashi:2018oca} with a model dependent upper limit that can be up to 1.2 TeV \cite{BarShalom:2012ms}. For this work we will therefore consider heavy neutrinos with masses of 45 GeV or larger. So our aim here is two-fold. First, we will show that the (Drell-Yan) production cross section for the isodoublet, driven by its EDM, is significantly larger than for an isosinglet. And second, that this  provides  a distinct advantage for it to be discovered using the MoEDAL detector, due to its lower luminosity  and being more forward peaked, as discussed  further on.

This paper will be organized as follows, in Section II we briefly describe MoEDAL's MAPP detector. In Section III, we will discuss the {\tt MadGraph} model we constructed in order to explore the potential detection of heavy neutrinos with large EDMs,  using MAPP.  In Section IV we briefly discuss the possibility that MoEDAL's  MAPP detector  can differentiate between a heavy neutral particle with large EDM and a mini-charged particle based on angular distribution.  In Section V we discuss our preliminary simulation of the detection of the neutrino EDM in the MAPP detector and show a plot that presents our sensitivity to heavy neutrinos, as described in the model presented here, with a detectable EDM. Finally we conclude in Section VI.   

\section{MoEDAL's MAPP Detector}
 \label{Detector}

In the work described here we utilize the central core of the MAPP detector that is designed to search for mini-charged particles (MAPP-mCP).  MAPP's other capability, the ability to  search for new long-lived weakly interacting neutral particles, is not relevant for this study. MAPP is protected from interacting Standard Model (SM) particles  at IP8 by roughly 25 m to 30 m of rock and from cosmic rays by an overburden of approximately 100 m of limestone. The MAPP detector can be deployed in a number of positions ranging from 5$^{\circ}$ to the beam  at at distance of $\sim$55 m from IP8  to approximately 30$^{\circ}$ to the beam at a distance of $\sim$ 5 m from IP8.  In this case we consider the small angle (5$^{\circ}$) position.

\begin{figure}[hbt]
	\begin{center}
 		\includegraphics[width=8cm]{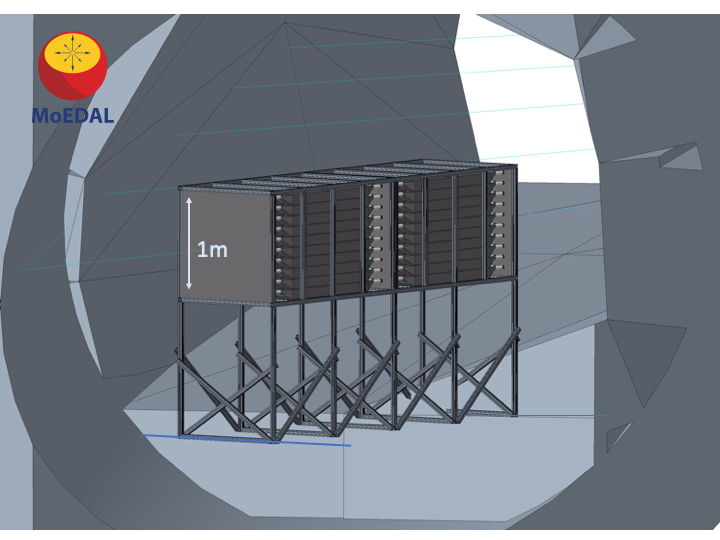}
 		\caption{Sketch of the  MAPP-mCP  subdetector. }\label{MAPP}
 	\end{center}
\end{figure}

The compact central section of MAPP that forms MAPP-mCP  is made up of two collinear sections, with cross-sectional area of 1.0 m$^{2}$, each comprised of  2 $\times$ 100 (10 cm $\times$ 10 cm) plastic scintillator bars each 0.75 m long. Thus, each through-going particle from the IP will encounter 3.0 m (4  $\times$ 75 cm)  of scintillator. Each bar is readout by a single low noise PMT.   All four PMTs are placed in coincidence in order to essentially eliminate backgrounds from dark counts in the PMTs and   radiogenic signals in the plastic scintillator or PMTs. The detectors are protected from cosmic rays and from particle interactions in the surrounding rock by charged particle veto detectors. A sketch of the MAPP-mCP detector is shown in Fig.~\ref{MAPP}.

\section{Model Validation and Production}
\label{Model}

As mentioned above there are several possible models that could possibly give rise to large EDMs. To the SM lepton representations, we add one vector-like doublet and its mirror. The vector-like doublet lepton will have both left-handed and right-handed components. The new representations are, with quantum number assignments for $SU(3)_c \times SU(2)_L \times U(1)_R$ given in brackets
\begin{eqnarray}
L_4= \left ( \begin{array}{c} N \\ E_4 \end{array} \right) =\left (1,2, -\frac{1}{2}\right),\  L_{c4}&=& \left ( \begin{array}{c} N_c \\ E_{c4} \end{array}\right) =\left (1,2, -\frac{1}{2} \right)\, ,
\end{eqnarray}
where we will assume that the mirror  doublet $L_{c4}$, which can also have interactions with the $Z$ boson,  is much heavier. We are particularly interested in the interaction of neutrinos in the model.   
 In order to cast a wide net in modelling the heavy neutrinos we use the following effective Lagrangian,
\begin{eqnarray}
& {\cal L}_{N}={\bar N}\left( i\slashed{\partial}-M_N\right) N+ie\ D{\bar N}\sigma_{\mu\nu}\gamma_5NF^{\mu\nu}\label{LagN}\\
&+ie D\tan\theta_W{\bar N}\sigma_{\mu\nu}\gamma_5NZ^{\mu\nu} +\displaystyle \frac{e}{2\cos\theta_W\sin\theta_W} Z^0_\mu{\bar N}_L\gamma^\mu N_L\nonumber
\end{eqnarray}
where the non-SM heavy neutrino is described by the field $N$, $A^\mu$ and $Z^{0\mu}$ denote the photon and $Z^0$ gauge fields, respectively, $F^{\mu\nu}=\partial^\mu A^\nu-\partial^\nu A^\mu$, and $Z^{\mu\nu}=\partial^\mu Z^{0\nu}-\partial^\nu Z^{0\mu}$. Here  $e D$ is the magnitude of the electron EDM,   and $M_N$ is the mass of the heavy neutrino.  The second and third terms of Eq. (\ref{LagN}) are effective low-energy dimension-five operators which involve the heavy neutrino $N$, seen as a massive neutral Dirac fermion, whose EDM (described in Eq. (\ref{LagN})), $e D$,  could be as large as $10^{-15}~e$ cm. This effective Lagrangian approach was pioneered  by Sher {\it et. al.} in Refs. \cite{Sher:2001rk} - \cite{Sher:2017wya}. Of particular interest to MoEDAL is  Ref.  \cite{Sher:2017wya} which discusses the search for heavy neutrinos with detectable EDMs at the LHC. 

In Ref. \cite{Sher:2017wya}, Sher and Stevens only considered heavy neutrino anti-neutrino production from quark-antiquark collisions interacting through an $s$-channel photon. This would correspond to using only the second term in Eq. \eqref{LagN}. Their neutrino is an isosinglet, has zero hypercharge and thus does not couple to the $Z$ boson. The isodoublet neutrino in our model couples to the $Z$, leading to a substantially higher production cross section.
In order to push beyond this we used the {\tt FeynRules} Mathematica package (\cite{Alloul:2013bka}) to implement our model in {\tt MadGraph} (\cite{Alwall:2011uj,Alwall:2014hca}), a matrix element evaluation tool.

\begin{figure}[hbt]
 	\begin{center}
 \includegraphics[width=6.5cm]{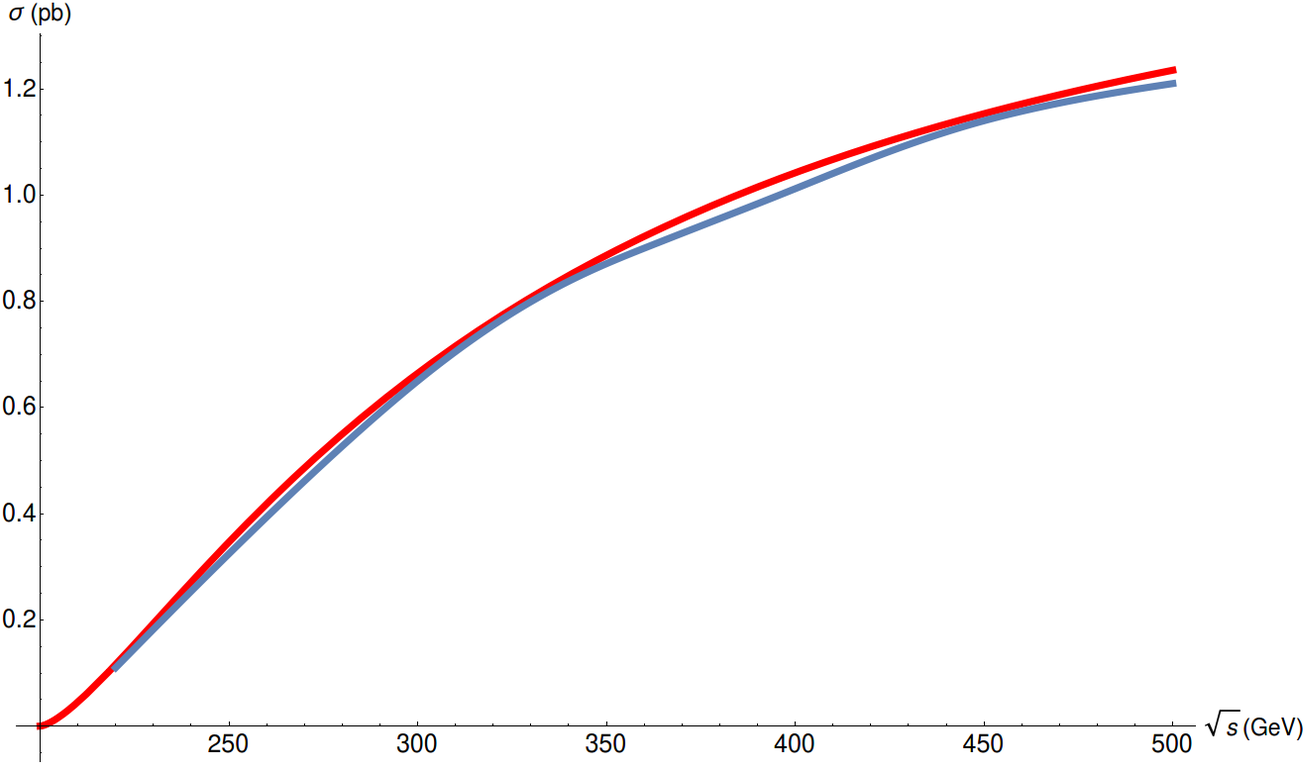} 		 \includegraphics[width=6.5cm]{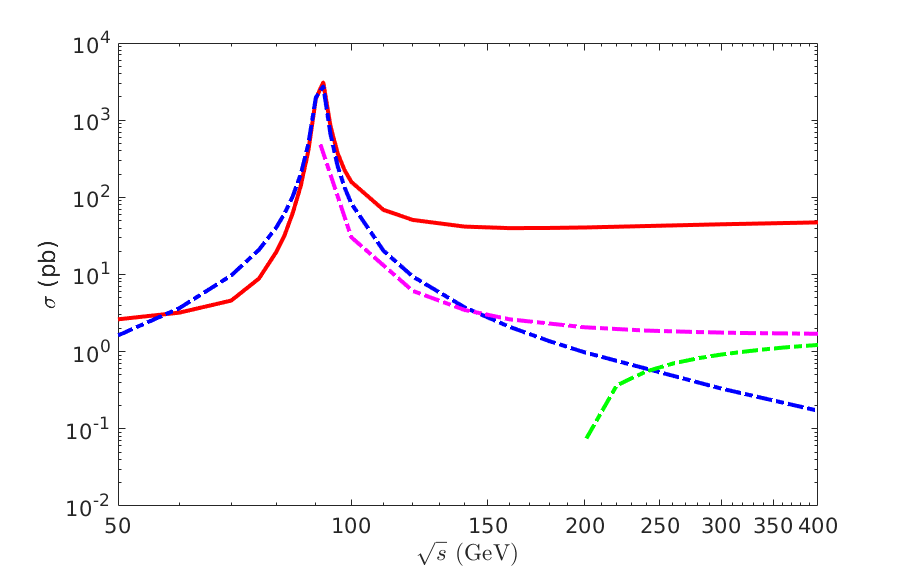}
 	\caption{Left: Comparison of $e^{+} \; e^{-}$ production cross section in {\tt MadGraph} (blue) with analytic expressions (red). Right: Comparison of $e^{+} \; e^{-} \; \rightarrow \bar{N} \; N $ (dashed) with $e^{+} \; e^{-} \; \rightarrow \bar{\nu}_{e} \; \ \nu_{e}$ (solid).  The blue, magenta, and green curves correspond to $M_{N}$ = 10, 45, and 100 GeV respectively.}\label{ZPole}	
 	\end{center}
 \end{figure}

To validate our model implementation we first looked at $e^{+} \; e^{-} \rightarrow N \; \bar{N}$ and considered only s-channel photons. This is effectively equivalent to only using  
\begin{equation}
{\cal L}_{\rm interaction \; 1}=ie D {\bar N}\sigma_{\mu\nu}\gamma_5NF^{\mu\nu}
\end{equation}
as the interaction term in our Lagrangian. Computing the differential cross section for this process gives
\[
\left(\frac{d\sigma}{d\Omega}\right)_\gamma=\frac14{\alpha^2 D^2}\left(1-\frac{4M_N^2}s\right)\sqrt{1-\frac{4M_N^2}s}\sin^2\theta\, ,
\]
where $\theta$ is the angle of the particle to the beam axis, from which a total cross section is easily determined.  Note that the term between our parentheses differs by a sign from the cross section in Eq. (2) of Ref. \cite{Sher:2001rk}.   After turning off the $Z$ contributions in the model, comparing this exact expression (shown as the red line) to the {\tt MadGraph} output (shown as the blue line on the plot) for our implementation gives excellent agreement as can be seen in the first graph of Fig.~\ref{ZPole}.

The model that we are using also stipulates that our heavy neutrino is a member of an isodoublet and will have the same interaction with the $Z$ as a regular neutrino would. In order to verify that our {\tt MadGraph} model correctly implemented this interaction we ``turned off'' the other interaction terms in the model and focused on
\begin{equation}
{\cal L}_{\rm interaction \; 2}=\frac e{2\cos\theta_W\sin\theta_W} Z^0_\mu{\bar N}_L\gamma^\mu N_L
\end{equation}
 
We then compared the cross section for $e^{+} \; e^{-} \; \rightarrow \bar{N} \; N $  to the cross section for production of a regular electron neutrino anti-neutrino pair in {\tt MadGraph}. The results of this can be seen in the second graph of Fig.~\ref{ZPole}. Note that both cross sections exhibit the characteristic $Z$ pole, but that the tail of the distribution is quite different due to the $N$'s substantially larger mass. Setting the $N$ mass to zero gives two identical distributions.

Note that we separately validated these two terms, one with a comparison to an analytic calculation and the second with a comparison to {\tt MadGraph}'s implementation of the standard model. The first term was validated by comparing to analytic calculation because standard {\tt MadGraph} does not contain any interactions of the form of the dimension 5 term that models the heavy neutrino's EDM. {\tt MadGraph} does, on the other hand, implement neutrinos so the second term was validated by comparing the new model to standard {\tt MadGraph}. Given that the two terms were shown to work separately we felt that there was no reason to validate them together with a further comparison to a much more complex analytic calculation.

Also, we looked at implementing a standard Yukawa interaction Higgs coupling for the heavy neutrino but as expected this did not substantially change the cross section for the heavy neutrino mass ranges in which we are interested.

\begin{figure}[hbt]
 	\begin{center}
 		 \includegraphics[width=8cm]{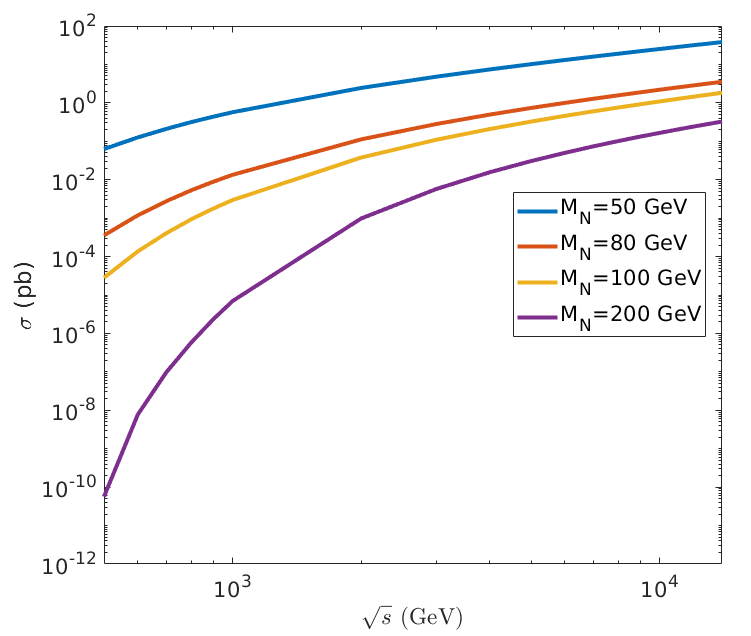}
 	\caption{Cross section for the production of $N$ $\bar{N}$ in a Drell-Yan process for neutrino masses of 50 GeV, 80 GeV, 100 GeV and 200 GeV.}\label{CrossSection3}	
 	\end{center}
 \end{figure}

We then considered the production of $N-\bar{N}$ via a Drell-Yan process using our model in {\tt MadGraph}. Fig.~\ref{CrossSection3} shows the cross section for this process as a function of centre of mass energy for several different heavy neutrino masses. 

\section{Angular Distribution on Heavy Neutrino Production}
  \label{Comparison}

In the model presented by  Sher and Stevens (\cite{Sher:2017wya}) the contribution to the differential cross section coming from EDM varies with $\sin^{2}(\theta)$,   which differs from that of mini-charged particles (mCP) which have a typical distribution of $1+\cos^{2}(\theta)$.  It might therefore be possible for MoEDAL to differentiate between this class of models and a more conventional mini-charged object.

\begin{figure}[hbt]
 	\begin{center}
 		 \includegraphics[width=9cm]{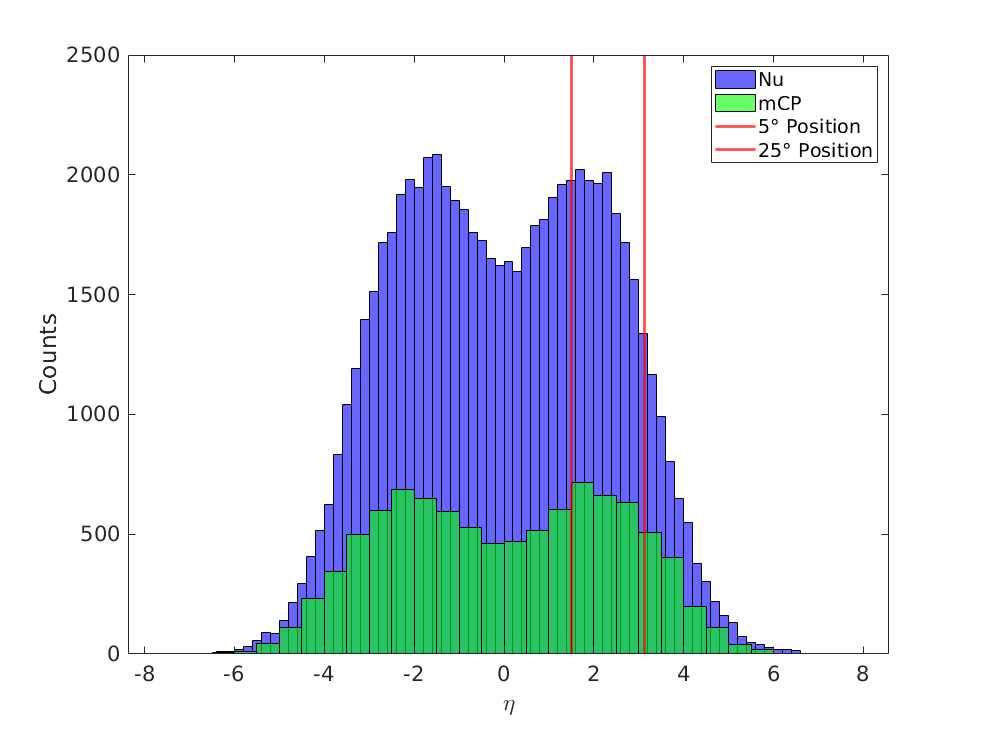}
 	\caption{Pseudorapidity distribution for events with a heavy neutrino with mass $M_N=100$ GeV and an EDM value $e D=10^{-16}\ e$ cm. \label{PseudoR} compared to the pseudopraidity distribution for mini-charged particles resulting from dark photon decays produced according to the scenario
 	explored in Ref. \cite{Haas:2014dda}.}
 	\end{center}
 \end{figure}

However, in the model we describe here, the heavy neutrinos are mostly forward-backward produced (typical to the Drell-Yan process) and in this case inclusion in the model of the weak interactions is highly relevant, as illustrated in Fig.~\ref{PseudoR}.  The angular distribution expected from mini-charged particles that arise in dark QED \cite{Holdom:1985ag}, for example the scenario explored in Ref. \cite{Haas:2014dda}, is also shown. Comparing these two models, one can see that the pseudorapidity distribution of heavy neutrino production falls of more  rapidly at high absolute  pseudorapidity than does that of mini-charged particle production. Additionally, the number of events detectable would at any pseudorapidity would be greater for heavy neutrino production, especially at smaller absolute luminosities. 

The mini-charged section of the  MoEDAL-MAPP detector can be moved along the UGCI gallery, adjacent to the LHCb/MoEDAL intersection region (IP8), from its nominal position at 
 at 5$^\circ$ to the LHC beamline to 25$^\circ$. As can be seen from Fig.~\ref{PseudoR}, it should be possible to distinguish the two models considered here using the MoEDAL-MAPP detector.
 
\section{Ionization Loss Due to the  EDM of the Heavy Neutrino}
 \label{simulation}

A neutrino with a large EDM can lose energy in a detector through electromagnetic interaction, thus rendering its detection possible \cite{Sher:2001rk}. As discussed in  \cite{Sher:2001rk,Sher:2017wya}, the impulse $\Delta\vec p=\int e\vec E\ dt$ given by the heavy neutrino's EDM to an atomic electron electron depends on the dipole's orientation. The impulse is equal to:
\[
\frac{eD}{4\pi\epsilon_0}\frac{2}{vb^2},
\]
 if the dipole is in the plane perpendicular to the neutrino's motion, the impulse is zero if the dipole is parallel to the neutrino's direction of motion.  Here, $e D$ is the size of the neutrino EDM and $v, b$ are the velocity of the neutrino and impact parameter to the atomic electron, respectively. For many interactions, the net average impulse given to an electron is expected to be half of this result. For a non-relativistic electron, this impulse leads to an energy transfer equal to
\[
\Delta E=\frac{\left|\Delta\vec p\right|^2}{2m}=\frac{e^4D^2}{2m\left(4\pi\epsilon_0\right)^2\left(vb^2\right)^2}
\]
With
\[
b_{\rm min}^2=\frac{e^2D}{2m\gamma v^2\left(4\pi\epsilon_0\right)},
\]
performing the integration cylindrically over the impact parameter, as in \cite{Sher:2001rk,Sher:2017wya}, we find:
\begin{equation}\label{BBF}
\frac{dE}{dx}=2\pi NZ\int_{b_{\rm min}}^\infty\ \Delta E(b)b\ db=\pi NZ\frac{e^2}{4\pi\epsilon_0}D\gamma,    
\end{equation}
where $Z$ is the nuclear charge, $N$ is the neutron number, and $\gamma=\frac 1{\sqrt{1-\beta^2}}$ the relativistic factor.

Detection of such heavy neutrinos will therefore depend on the size of its EDM and its mass. Using our {\tt MadGraph} model with $\sqrt{s}=14$ TeV we generate Drell-Yan produced heavy neutrinos. Using Eq.~\ref{BBF}, we then simulate their energy loss through 25 m of rock,  the average amount of material would be encountered by a neutrino impinging on the MoEDAL-MAPP detector deployed at 5$^{\circ}$ to the beam line,  followed by an air-gap and then 3 m of plastic scintillator. We assume that the heavy neutrino would be detected due to its EDM if it gives rise to 100 photons or more in each of the 4 sections of the detector for a total of at least 400 photons. To convert energy deposition into number of photons in the scintillator we assume that $10^{4}$ photons are produced per MeV of energy deposited in the  plastic scintillator \cite{Tanabashi:2018oca}.

\begin{figure}[hbt]
 	\begin{center}
  		 \includegraphics[width=8cm]{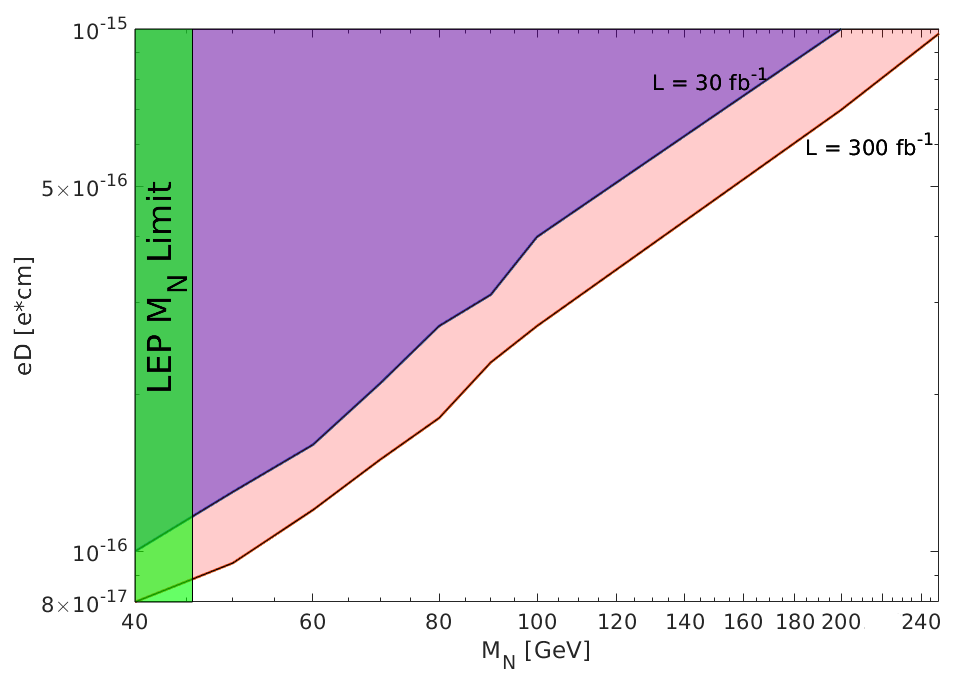}
 	\caption{The reach for heavy neutrino EDM detection at MoEDAL's MAPP detector at $\sqrt{s}=14$ TeV, with 3 or more events observed at 95\% C.L., and 30 ${\rm fb^{-1}}$ and 300 ${\rm fb^{-1}}$ of integrated luminosity.}\label{Exclusion}	
 	\end{center}
 \end{figure}

For ease of comparison with our detectors, we assume that the MAPP detector is 100\% efficient.  Given this assumption, our `best case' sensitivity contour to heavy neutrino
EDM observation is indicated by 3 or more events observed at 95\% C.L. for each value of $e D$ and $M_{N}$, in  Fig. \ref{Exclusion}. In this plot we considered both 30 ${\rm fb^{-1}}$ (LHC's Run-3)  and 300 ${\rm fb^{-1}}$ (High Luminosity LHC)  of integrated luminosity taken at IP8.  We are currently studying the response of the detector to backgrounds such as: neutrons and  KoLs from collisions at IP8 that penetrate the rock shielding in front of MAPP;  and,  also from cosmic ray interactions in the rock surrounding the MAPP detector. Concomitantly, we are studying the role of timing, tracking and pointing in the reduction of these potential backgrounds.

We see that with 30 ${\rm fb^{-1}}$ of data available to MoEDAL during Run-3 of the LHC, MAPP will be able to exclude heavy neutrino masses from 40-200 GeV with EDM values as low as $10^{-16} \ e$ cm in the most favorable scenario.  Tighter bounds predicted assuming 300 fb$^{-1}$ of data improve our reach slightly, down to ~$8 \times 10^{-17} \ e$ cm.

\section{Conclusion} 
\label{Section4}

In our work we have extended the work in ~\cite{Sher:2017wya} and implemented a heavy neutrino model where the heavy neutrino is a member of an isodoublet. We have further considered this in the context of MoEDAL's MAPP detector and concluded,  based on these initial studies, that a heavy neutrino  with a large enough  EDM could in principle be detected at at the LHC using MoEDAL's MAPP detector. A non-observation would allow us  place bounds on the value of $D$ as well as on the mass of such particles.




\section*{Acknowledgement}

We are grateful to the Natural Sciences and Engineering Research Council (NSERC) of Canada for partial financial support (grant number sAPPJ-2019-00040).  M.F. thanks NSERC for partial financial support under grant number SAP105354. M. de Montigny acknowledges NSERC for partial financial support (grant number RGPIN-2016-04309). We would also like like to thank Mo Abdullah for his valuable input.





\end{document}